\documentclass[journal]{IEEEtran}
\usepackage{cite}
\usepackage{bm}
\ifCLASSINFOpdf
 \usepackage{graphicx}
   \DeclareGraphicsExtensions{.pdf,.jpeg,.png,.eps}
\else
  \usepackage[dvips]{graphicx}
\fi
\usepackage[cmex10]{amsmath}
\usepackage{array}
\usepackage{url}

\begin{document}
%
\title{Analysis of Metal-Dielectric Waveguides with Circular Sectors}
%
%
%

\author{Daniel~Torrent,~Benito~Gimeno,~Vicente E. Boria~and~Jos\'e S\'anchez-Dehesa}      

%
%

\markboth{Journal of \LaTeX\ Class Files,~Vol.~6, No.~1, January~2007}%
{Shell \MakeLowercase{\textit{et al.}}: Bare Demo of IEEEtran.cls for Journals}
%



\maketitle

\begin{abstract}
The study of metallic corrugated surfaces has recently received strong attention due to their ability to mimic the behaviour of surface plasmons. In this work, this plasmon-like behaviour is employed to design an open cylindrical waveguide. The structure consists on a longitudinally corrugated metallic cylinder with corrugations filled with a dielectric material. The dispersion relation of this waveguide is analyzed in the Òcontinuous limitÓ, defined as the limit in which the number of corrugations is infinite, but keeping their periodicity constant. It is found that, in this limit, the waveguide supports only TE guided modes and their dispersion relation becomes highly degenerated. Finally, it is shown that the waveguide behaves as an anisotropic cylindrical rod with extreme electromagnetic parameters, what makes it possible to apply these structures not only as waveguides but also as building blocks for metamaterials.\end{abstract}

\begin{IEEEkeywords}
IEEEtran, journal, \LaTeX, paper, template.
\end{IEEEkeywords}

%
\IEEEpeerreviewmaketitle

\section{Introduction}
\IEEEPARstart{P}{eriodic} corrugated dielectric and/or metallic surfaces have been widely studied for their use as waveguide structures due to their relevant properties
\cite{Collin1991}. Periodicity impinges additional control over the propagation characteristics of a waveguide structure, which otherwise are limited to the properties of the dielectric or metallic materials employed in their design. 

Recently, it has been shown that periodically corrugated surfaces can be used to mimic plasmons in the microwave or terahertz 
frequency bands \cite{Pendry2004, Hibbins2005,GarciadeAbajo2005,GarciadeAbajo2007,Kats2011}. Furthermore, it has been shown that texturing closed cylindrical or spherical surfaces allows the localization of these plasmon-like waves, showing as well as that such closed surfaces behave as an 
 anisotropic material with extreme values of its permitivity and permeability\cite{Pors2012}. This effective behavior had already been studied in the past for the definition and  realization of soft and hard electromagnetic boundaries \cite{Kildal1990}.

In this context, the control of the effective electromagnetic properties by means of artificially designed structures has recently received strong attention. By naming 
metamaterials to the resulting effective material, a wide variety of new and exciting phenomena and applications have been found, and several methods of metamaterials realization have been also considered. In this area special mention deserves the author's  work concerning 
the realization of electromagnetic metamaterials using highly anisotropic cylinders\cite{Carbonell2011,TorrentNJP11,Carbonell2012,Carbonell2013}.

In this letter we analyze a cylindrical waveguide structure based on the localization of plasmons placed in a lossless open circular environment. The surface consists on a 
longitudinally corrugated metallic cylinder where the grooves are filled with a dielectric material, as shown in Fig.\ref{schematics}.
It will be demonstrated that the waveguide allows only the propagation of TE modes; moreover, the dispersion relation of the different multipolar components are 
shown to be quite similar, being identical under some special conditions explained in this work. Finally, the extraordinary anisotropic electromagnetic properties
of the novel proposed waveguide are given, thus providing the potential applications of these cylindrical structures for the realization of electromagnetic metamaterials.

\section{Theory}

\begin{figure}
\centering
\includegraphics[width=\columnwidth]{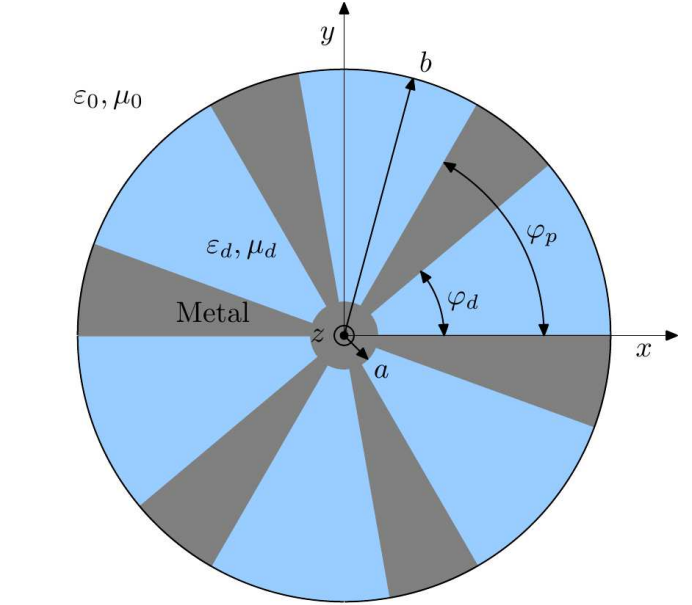}
\caption{ \label{schematics}Geometry of the angularly layered waveguide.}
\end{figure}
The geometry of the proposed uniform waveguide is an inhomogeneous cylindrical waveguide, as depicted in Fig. \ref{schematics}. An infinitely long cylinder of 
radius $b$ with its axis parallel to the $z$ axis is considered. The cross section of the cylinder is made of alternating sectors of a perfect conductor combined with a dielectric material of permittivity $\varepsilon_d$ and magnetic permeability $\mu_d$, called the ``propagation material''; the structure is immersed in free space, characterized by $\mu_0$ and $\varepsilon_0$. A metallic core of radius $a$ is placed at the center of the cylinder. The waveguide is divided into $N$ identical sectors of angle
$\varphi_p=2\pi/N$, being each sector divided into two subsectors, one filled with the propagation material having an angle $\varphi_d$, and the other one filled with the perfect conductor material having an angle $\varphi_m=\varphi_p-\varphi_d$. The angular geometry of the waveguide is defined such that the propagation material sectors start
at $\varphi_d^{(n)}=(n-1)\varphi_p$, while the conductor sectors start at $\varphi_m^{(n)}=(n-1)\varphi_p+\varphi_d$, with $n=1,2,\dots,N$.
\par
Considering a time-harmonic dependence with frequency $f=\omega/(2\pi)$ and invariance along the $z$ axis, it is known that the transverse components of the electromagnetic guided fields can be obtained from the $z$ components as follows \cite{Jackson1998}
\begin{subequations}
\label{eq:EtHt}
\begin{eqnarray}
\bm{E}_t &=&\frac{-j}{\mu\varepsilon\omega^2-\beta^2}\left[ \beta\bm{\nabla}_t E_z +\omega\mu\bm{\hat{z}}\times\bm{\nabla}_t H_z\right]\\
\bm{H}_t &=&\frac{-j}{\mu\varepsilon\omega^2-\beta^2}\left[ \beta\bm{\nabla}_t H_z -\omega\varepsilon\bm{\hat{z}}\times\bm{\nabla}_t E_z\right]
\end{eqnarray}
\end{subequations}
where $\beta$ is the modal propagation constant, and $\mu$ and $\varepsilon$ represent the magnetic permeability and the dielectric permittivity, respectively; it should be emphasized that both $\mu$ and $\varepsilon$ are functions on the transverse vector position. Boundary equations, together with the guided wave condition, will be forced in order to obtain the electromagnetic fields for $r<b$ and for $r>b$, while
at $r=b$ mode matching method will grant the satisfaction of the corresponding boundary conditions. For $r>b$ only evanescent solutions are allowed, thus the $z$ components of the fields will be expressed as
\begin{subequations}
\label{eq:rga}
\begin{eqnarray}
E_z(r,\varphi)=\sum_{q=-\infty}^{\infty}A_q^E K_q(\Gamma r)e^{jq\varphi},\\
H_z(r,\varphi)=\sum_{q=-\infty}^{\infty}A_q^H K_q(\Gamma r)e^{jq\varphi},
\end{eqnarray}
\end{subequations}
where $K_q(\cdot)$ are modified Bessel functions of second kind, which are evanescent-like solutions which cancels as $r\to\infty$, and
$\Gamma^2=\beta^2-\mu_0\varepsilon_0\omega^2$, $\Gamma$ being the propagation constant for $r>b$.
\par
Inside the waveguide, that is, for $r<b$, the fields must satisfy the proper boundary conditions at the walls of the metallic interface\cite{Jackson1998}, that is,
at $\varphi=\varphi_d^{(n)}$ and $\varphi=\varphi_m^{(n)}$. A proper solution for the fields is now a linear combination of the modes of a circular sector 
cavity. Thus, for $\varphi\in[\varphi_d^{(n)},\varphi_m^{(n)}]$ we have
\begin{subequations}
\label{eq:rla}
\begin{eqnarray}
E_z(r,\varphi)=\sum_{s=1}^{\infty}B_{ns}^E J_{s\gamma}(k_{td} r)\sin s\gamma(\varphi-\varphi_d^{(n)}) ,\\
H_z(r,\varphi)=\sum_{s=0}^{\infty}B_{ns}^HJ_{s\gamma}(k_{td} r)\cos s\gamma(\varphi-\varphi_d^{(n)}) , \label{eq:rlab}
\end{eqnarray}
\end{subequations}
$ k_{td}^2=\mu_d\varepsilon_d\omega^2-\beta^2$ being the transversal wavenumber inside the propagating material, and $J_{s\gamma}(\cdot)$ are Bessel functions of first kind and real order $\gamma s$ with
\begin{equation}
\label{eq:gamma}
\gamma=\frac{\pi}{\varphi_d}.
\end{equation}
\par 
A metallic circular boundary of radius $a$ placed inside the waveguide will be required in a practical application, which can be easily considered into the equations 
through the following replacements
\begin{subequations}
\begin{equation}
J_{s\gamma}(k_{td} r)\rightarrow J_{s\gamma}(k_{td} r)-\frac{J_{s\gamma}(k_{td} a)}{Y_{s\gamma}(k_{td} a)}Y_{s\gamma}(k_{td} r)
\end{equation}
for the $E_z$ field
and
\begin{equation}
J_{s\gamma}(k_{td} r)\rightarrow J_{s\gamma}(k_{td} r)-\frac{J'_{s\gamma}(k_{td} a)}{Y'_{s\gamma}(k_{td} a)}Y_{s\gamma}(k_{td} r)
\end{equation}
for the $H_z$ field. The objective of this work is to study this type of waveguide in its simplest form,  thus it will be assumed that $a=0$, since finite but small values of 
$a$ only contribute to the dispersion relation.
\end{subequations}
\par
Once the solution for the fields at both sides of the boundary $r=b$ is established, standard mode matching method is directly applied. Boundary conditions
at $r=b$ implies the continuity of the $z$ and $\varphi$ components of the $\bm{E}$ and $\bm{H}$ fields at the boundaries of the propagating material sector, as well as the cancellation of
$E_z$ and $E_\varphi$ at the corresponding boundaries of the perfect metallic sectors. In order to apply the mode-matching technique, the equations for $E_z$ and $E_\varphi$ must be multiplied by the modal solutions outside the waveguide, while the
equations for the $H_z$ and $H_\varphi$ must be multiplied by the solution inside the waveguide. After some algebraic manipulations, equations are integrated within the corresponding angular regions so that we can 
obtain the unknown coefficients $A_q^E, A_q^H,B_{ns}^E$ and $B_{ns}^H$ defined in equations \eqref{eq:rga} and \eqref{eq:rla}.  For instance, the continuity condition of the $E_z$ component is imposed after multiplying by the exponential factor $e^{-jq\varphi}$ and integrated over $\varphi \in [0,2\pi]$,
\begin{subequations}
\label{eq:system}
\begin{equation}
\label{eq:mez}
A_q^EK_q(\Gamma b)=\frac{1}{2\pi }\sum_{n=1}^N\sum_{s=1}^{\infty}B_{ns}^EJ_{s\gamma}(k_{td} b)N_{sq}^{(n)*},
\end{equation}
where $*$ means complex conjugate operation, and $N$ is the total number of angular sectors. Next, the axial magnetic field component is multiplied by $\cos s\gamma(\varphi-\varphi_d^{(n)})$, and integrated 
over $\varphi \in [\varphi_d^{(n-1)},\varphi_d^{(n)}]$, then we obtain for the $n$-th angular sector the following relationship
\begin{equation}
\label{eq:mhz}
\sum_{q=-\infty}^{\infty}A_q^HK_q(\Gamma b)M_{sq}^{(n)}=(1+\delta_{0s})\frac{\pi}{2\gamma}B_{ns}^HJ_{s\gamma}(k_{td} b).
\end{equation}
where $\delta_{0s}$ is the Kronecker delta, and  the coupling matrix-elements are given by
\begin{IEEEeqnarray*}{l}
M_{sq}^{(n)}\equiv\int_{\varphi_d^{(n)}}^{\varphi_m^{(n)}}\cos s\gamma(\varphi-\varphi_d^{(n)}) e^{jq\varphi}d\varphi=\nonumber\\
\frac{\pi}{2\gamma}e^{jq\varphi_d^{(n)}}\left[e^{i\alpha_+}\sin\alpha_+/\alpha_++e^{i\alpha_-}\sin\alpha_-/\alpha_- \right]\\
N_{sq}^{(n)}\equiv\int_{\varphi_d^{(n)}}^{\varphi_m^{(n)}}\sin s\gamma(\varphi-\varphi_d^{(n)}) e^{jq\varphi}d\varphi=\nonumber\\
-\frac{j\pi}{2\gamma}e^{jq\varphi_d^{(n)}}\left[e^{i\alpha_+}\sin\alpha_+/\alpha_+-e^{i\alpha_-}\sin\alpha_-/\alpha_- \right]
\end{IEEEeqnarray*}
being $\alpha_\pm=\frac{\pi}{2}[(q/\gamma)\pm s]$. Also, after some algebra, the equation for the $E_\varphi$ component is multiplied
by $e^{-jq\varphi}$ and integrated over $\varphi \in [0,2\pi]$, 
\begin{IEEEeqnarray}{l}
\label{eq:mephi}
\frac{jq\beta}{\Gamma^2a}K_q(\Gamma b)A_q^E-\frac{\omega\mu_0}{\Gamma}K_q'(\Gamma b)A_q^H=\nonumber\\
-\frac{1}{2\pi }\sum_{n=1}^N\sum_{s=0}^\infty M_{sq}^{(n)*}\times\nonumber\\
\left[ \frac{s\gamma\beta}{k_{td}^2a}J_{s\gamma}(k_{td} b)B_{ns}^E-\frac{\omega\mu_d}{k_{td}}J_{s\gamma}'(k_{td} b)B_{ns}^H\right].
\end{IEEEeqnarray}
 Finally, the equation for $H_\varphi$ is multiplied by $\sin s\gamma(\varphi-\varphi_d^{(n)})$, and integrated 
over $\varphi \in [\varphi_d^{(n-1)},\varphi_d^{(n)}]$, resulting in
\begin{IEEEeqnarray}{l}
\label{eq:mhphi}
\sum_{q=-\infty}^\infty\left[\frac{jq\beta}{\Gamma^2a}K_q(\Gamma b)A_q^H+\frac{\omega\varepsilon_0}{\Gamma}K_q'(\Gamma b)A_q^E\right]N_{sq}^{(n)}
=\nonumber\\
-\frac{\pi}{2\gamma}\left[ \frac{s\gamma\beta}{k_{td}^2a}J_{s\gamma}(k_{td} b)B_{ns}^H+\frac{\omega\varepsilon_d}{k_{td}}J_{s\gamma}'(k_{td} b)B_{ns}^E\right].
\end{IEEEeqnarray}
\end{subequations}
Equations \eqref{eq:system} define a homogeneous system of equations for the unknown coefficients $A_q^E, A_q^H, B_{ns}^E$ and $B_{ns}^H$. The frequency
values that make the determinant of the corresponding matrix equation vanishes for positive values of $\Gamma^2$ define the guided modes. The discussion on the several 
solutions of this system is out of the scope of the present work. Here we are mainly interested in the analysis of a limiting case where equations simplify, which 
corresponds to a type of waveguides whose properties can be easily understood.
\section{Continuous Limit}
 Let us define the filling fraction $\xi$ as the ratio of the area occupied by the propagation material divided by the sector area of the waveguide,
 \begin{equation}
\xi\equiv\frac{\varphi_d}{\varphi_p}=\frac{N}{2\gamma},
\end{equation}
where \eqref{eq:gamma} has been employed.
\par
The continuous limit is defined as the limit when the number of cells goes to infinity, $N\to\infty$, while the filling fraction $\xi$ remains constant, so that we also have 
that
$\gamma\to\infty$, which is obvious as in this limit $\varphi^{(n)}_d\to\varphi^{(n)}_m$. In this limit, the structure will behave as a continuous and homogeneous material, as it will be demonstrated next in this section.
\par
In the continuous limit, it is easy to see that the matrix elements $N_{sq}^{(n)}$ cancels, therefore from \eqref{eq:mez} we
easily obtain
\begin{subequations}
\begin{equation}
\label{eq:Ez0}
A_q^EK_q(\Gamma b)=0 \Rightarrow A_q^E=0.
\end{equation}
This result shows that this structure only allows the propagation of $TE$ modes.
In the same limit, the Bessel functions $J_{s\gamma}(\cdot)$ are identically zero for all value of $s$ but $s=0$, thus again it is easy to see that equation \eqref{eq:mhz} becomes
\begin{equation}
\label{eq:aqh1}
\sum_{q=-\infty}^{\infty}A_q^HK_q(\Gamma b)M_{sq}^{(n)}=\frac{\pi}{\gamma}B_{n0}^H\delta_{0s}J_0(k_{td}b),
\end{equation}
where we still keep the $\gamma$ factor in the right hand side because it will be cancelled by the $\gamma$ factor in the denominator of $M_{sq}^{(n)}$.
\par
Similarly, the reduced form of equation \eqref{eq:mephi}, reminding that $A_q^E=0$, can be cast as follows
\begin{equation}
\label{eq:aqh2}
-\frac{\omega\mu_0}{\Gamma}K_q'(\Gamma b)A_q^H=\frac{1}{2\pi }\sum_{n=1}^NM_{0q}^{(n)*}\frac{\omega\mu_d}{k_{td}}B_{n0}^HJ_0'(k_{td}b).
\end{equation}
\end{subequations}
Finally, from Eq. \eqref{eq:mhphi} we arrive to a trivial solution, because the left hand side is zero due to the cancellation of the matrix elements $N_{sq}^{(n)}$, 
while the right hand side is zero due to the terms involving $J_{s\gamma}$.
\par
Solving for $B_{n0}^H$ from \eqref{eq:aqh1}, and inserting it into \eqref{eq:aqh2}, we find
\begin{equation}
-\frac{\mu_0}{\Gamma}K_q'(\Gamma b)A_q^H=\frac{\mu_d}{k_{td}}\sum_{s=-\infty}^\infty\chi_{qs}A_s^HK_s(\Gamma b),
\end{equation}
where the $\chi_{qs}$ matrix elements are defined as
\begin{equation}
\chi_{qs}=\frac{J_0'(k_{td}b)}{J_0(k_{td}b)}\sum_{n=1}^N\frac{\gamma}{2\pi^2}M_{0q}^{(n)*}M_{0s}^{(n)},
\end{equation}
which still keeps finite values of both $N$ and $\gamma$. 
\par
Although the above system of equations could be used as an approximated solution
to \eqref{eq:system}, for large values of $N$ and $\gamma$, the goal of this work is to analyze the continuous limit. As a consequence, the infinite limit 
of both $N$ and $\gamma$ has to be taken, leading to
\begin{equation}
\lim_{N\to\infty}\chi_{qs}=\frac{J_0'(k_{td}b)}{J_0(k_{td}b)}\lim_{N\to\infty}\frac{\xi}{N}\sum_{n=1}^Ne^{i(s-q)\varphi_d^{(n)}}=\xi\frac{J_0'(k_{td}b)}{J_0(k_{td}b)}\delta_{qs}.
\end{equation}

This result demonstrates that in the continuous limit all the multipolar $q$ components are fully decoupled, and the dispersion relation simply becomes
as follows
\begin{equation}
\label{eq:dispersion}
\frac{\mu_0}{\Gamma}\frac{K_q'(\Gamma b)}{K_q(\Gamma b)}=\frac{\mu_d\xi}{k_{td}}\frac{J_1(k_{td}b)}{J_0(k_{td}b)}.
\end{equation}
This equation defines a transcendental equation to determine the dispersion relation $\beta=\beta(k_0)$ (where $k_0=\omega\sqrt{\varepsilon_0\mu_0}$ is the wavenumber in free space), for the continuous limit
of the sectorial waveguide previously analyzed in its general form. The solutions for $\beta=\beta(k_0)$ cannot be obtained analytically from \eqref{eq:dispersion}, and
a numerical root finding algorithm must be employed.
\par
Note that for the case $q=0$, $\mu_d=\mu_0$ and $\xi=1$, the eigenvalue equation for the $TE$ modes is recovered in \eqref{eq:dispersion}; however, for the case $q\neq 0$ the
eigenvalue equation is completely different even when $\xi=1$. In fact, even though the case $\xi=1$ indicates that the full waveguide
is made of dielectric material, the boundary condition due to the perfect metallic walls is still present in the equations and, as a consequence, the result turns in a completely different scenario. 
\par
It is interesting to note that all the modes resulting from \eqref{eq:dispersion} have the same cut-off frequency, corresponding to the zeros of Bessel function $J_0(k_{td}b)$. Effectively, 
the cut-off frequency is obtained in the limit $\beta\to k_0$ which corresponds to $\Gamma\to0$. In this
limit \eqref{eq:dispersion} can only be hold if $J_0(k_{td} b)=0$, which leads to 
\begin{equation}
k_{td} b=\sqrt{\mu_d\varepsilon_d \omega^2-\beta^2}b=\omega b\sqrt{\mu_d\varepsilon_d-\mu_0\varepsilon_0}=\kappa_i,
\end{equation}
where $\kappa_i$ are the zeros of the zero order Bessel function, i.e., $J_0(\kappa_i)=0$. Thus, the set of cut-off frequencies are given by
\begin{equation}
\omega_i b=\frac{\kappa_i}{\sqrt{\mu_d\varepsilon_d-\mu_0\varepsilon_0}}.
\end{equation}
It should be noted that these frequencies are independent of $q$ and $\xi$, being therefore only dependent of the material employed to construct the waveguide.
\par
Another interesting result is found when we take the limit $q\to\infty$, then we have that
\begin{equation}
\lim_{q\to\infty}\frac{\mu_0}{\Gamma}\frac{K_q'(\Gamma b)}{K_q(\Gamma b)}=\infty
\end{equation}
and the dispersion relation $\beta(k_0)$ is found from the condition $J_0(k_{td} b)=0$ or $k_{td} b=\kappa_i$, easily obtaining
\begin{equation}
\beta_\infty^{(i)}=\sqrt{\mu_d\varepsilon_d\omega^2-(\kappa_i/b)^2}.
\end{equation}
\par 
Next section solves \eqref{eq:dispersion} for a particular case of practical interest, and it will be shown that such equation can be
properly used to characterize this novel type of waveguide.
\section{Numerical Results}
Let us consider the case where $\varepsilon_d=2.56\,\varepsilon_0$ and $\mu_d=\mu_0$, 
which corresponds to the dielectric material used in \cite{Collin1991} to analyze the dielectric rod waveguide.
\par
\begin{figure}

\centering
\includegraphics[width=\columnwidth]{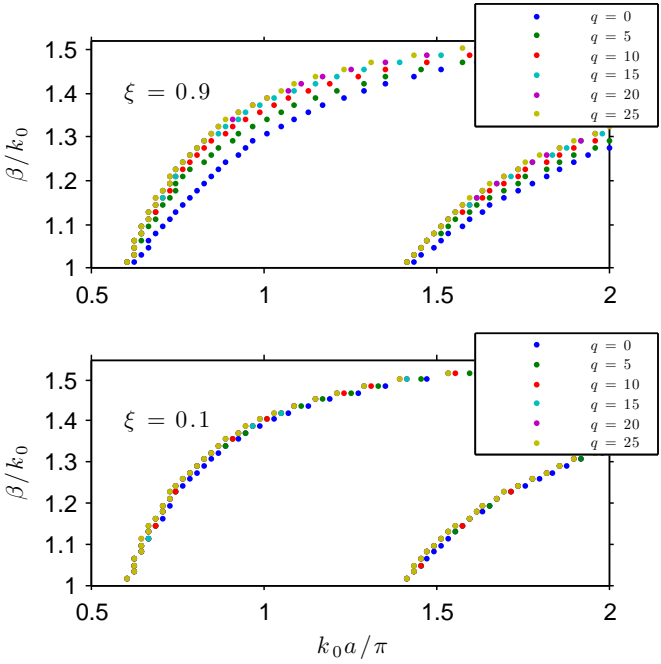}
\caption{ \label{figure_disp} Dispersion relationship for the angularly layered waveguide. }
\end{figure}
Fig. \ref{figure_disp} shows the dispersion relationship obtained from \eqref{eq:dispersion} for two filling fraction values of $\xi=0.9$ (upper side) and $\xi=0.1$ (lower 
side). The several depicted dots are the ones corresponding to the multipolar components $q=0$ to $q=25$ in steps of 5. 
Note also that even for the case of a low metallic filling fraction ($\xi=0.9$) the modal dispersion is small, being the case of $\xi=0.1$ of practically zero 
modal dispersion. Moreover, for the two cases shown in Fig. \ref{figure_disp}, the modal dispersion relationship for large values of $q$ are quite similar. 
Thus, the dispersion relation $\beta_\infty$ can be used as a reference for characterizing this type of waveguides, since from Figure \ref{figure_disp}
it can be concluded that all the modes present a similar dispersion behavior.
\par
\begin{figure}

\centering
\includegraphics[width=\columnwidth]{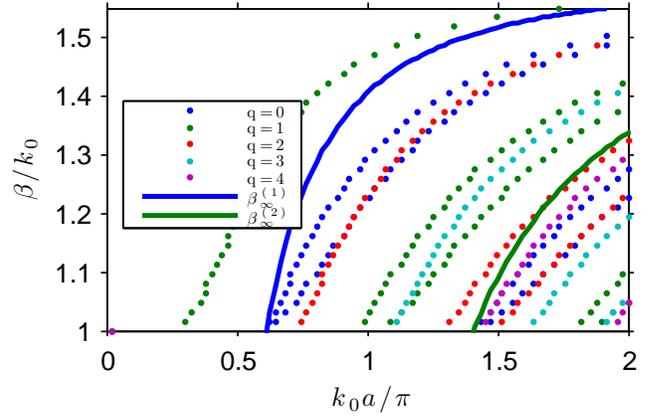}
\caption{\label{figure_optical} The dispersion relationship of a dielectric rod with $\varepsilon_d=2.56\,\varepsilon_0$ and $\mu_d=\mu_0$ (dots), is compared with the limiting dispersion relation $\beta_\infty/
k_0$ (continuous lines) of the angularly layered waveguide filled with the same dielectric material.
}
\end{figure}

 
Fig. \ref{figure_optical} shows the dispersion response $\beta_\infty^{(i)}$ for $i=1,2$ (continuous lines) together with the dispersion curves for a dielectric
rod (dotted line) made of the same dielectric material ($\varepsilon_d=2.56\,\varepsilon_0$ and $\mu_d=\mu_0$). Note that the modal dispersion is larger for the dielectric rod, where the single-mode regime is much smaller than for the angularly layered waveguide, as it is reported in \cite{Collin1991}.
\par 
\section{Modal Solution and Effective Electromagnetic Tensor}
It has been already shown that the polarization of the EM fields traveling along the waveguide is TE, since $E_z=0$. Let us deduce the expressions for the
other EM-field components in the continuous limit defined in the previous section. The $H_z$ component is given by \eqref{eq:rlab}, while the $B_{ns}^H$ is
obtained from \eqref{eq:aqh1}, showing that this coefficient cancels for all $s$ but $s=0$, thus it can be shown that the $H_z$ field inside the cylinder,
for $\varphi\in[\varphi_d^{(n)},\varphi_m^{(n)}]$, is given by
\begin{equation}
H_z(r,\varphi)=\sum_{q=-\infty}^{\infty}A_q^H\frac{K_q(\Gamma b)}{J_0(k_{td}b)}J_0(k_{td}r)\frac{\gamma}{\pi}M_{0q}^{(n)}
\end{equation}
The factor $\gamma M_{0q}^{(n)}$ is, in the limit $\gamma\to\infty$, proportional to $e^{jq\varphi_d}\approx e^{jq\varphi}$, thus the $H_z$ component inside the
cylinder has the following form
\begin{equation}
\label{eq:solHz}
H_z(r,\varphi)=J_0(k_{td}r)\sum_{q=-\infty}^{\infty}B_q^He^{jq\varphi}
\end{equation}
which shows that the radial dependence of the field is completely decoupled from the angular dependence, as it might be deduced from \eqref{eq:dispersion}.

Additionally, from Eqs. \eqref{eq:EtHt} and \eqref{eq:rla}, together with the fact that $B_{ns}^H=B_{n0}^H\delta_{0s}$, we can conclude that the fields component $E_r$ and
$H_\varphi$, both implying terms proportional to the angular derivative of $H_z$, therefore to $s B_{ns}^H$, will vanish inside the cylinder, then
$E_r(r,\varphi)=0$ and $H_\varphi(r,\varphi)=0$.

The latter of these equations will be also a boundary condition, since the tangential component of the fields must be continuous at an arbitrary interface. Equations \eqref{eq:EtHt} 
show also that both $E_\varphi$ and $H_r$ will be different than zero and proportional to $\partial_rH_z$. 

In summary, the only nonzero components of the field within this structure, in the continuous limit, will be $H_z, H_r$ and $E_\varphi$. In this limit,
the expression for the $H_z$ field is given by \eqref{eq:solHz}, but the expressions for $H_r$ and $E_\varphi$ cannot be obtained
by a direct application of \eqref{eq:EtHt}, since these equations are not valid in the continuous limit.

In this limit, the cylindrical structure analyzed in the present work behaves as a homogeneous and anisotropic
cylinder, thus the relationship between the electromagnetic field components must satisfy the tensorial generalization of Eqs. \eqref{eq:EtHt}, obtained
from Maxwell equations in tensor form as indicated in \cite{Chew1999}
\begin{subequations}
\label{eq:aniMax}
\begin{eqnarray}
\nabla\times\bm{E}+j\omega\overline{\overline{\mu}}\cdot\bm H=0\\
\nabla\times\bm{H}-j\omega\overline{\overline{\varepsilon}}\cdot\bm E=0
\end{eqnarray}
\end{subequations}
where the tensors $\overline{\overline{\mu}}$ and $\overline{\overline{\varepsilon}}$ are diagonal matrices of the form
\begin{equation}
\overline{\overline{\varepsilon}}=\left(
\begin{array}{ccc}
\varepsilon_r & 0 & 0 \\
0 & \varepsilon_\varphi & 0 \\
0 & 0 &\varepsilon_z \\
\end{array} \right )=\left(\begin{array}{cc} \overline{\overline{\varepsilon}}_t & 0\\ 0 & \varepsilon_z\end{array}\right)
\end{equation}
and
\begin{equation}
\overline{\overline{\mu}}=
\left(
\begin{array}{ccc}
\mu_r & 0 & 0 \\
0 & \mu_\varphi& 0 \\
0 & 0 &\mu_z \\
\end{array} \right )=\left(\begin{array}{cc} \overline{\overline{\mu}}_t & 0\\ 0 & \mu\end{array}\right)
\end{equation}

After replacing $\bm\nabla$ by $\nabla_t-j\beta\bm\hat{z}$ in \eqref{eq:aniMax}, these equations can be cast as
\begin{subequations}
\begin{eqnarray}
-j\beta \bm E_t+j\omega \bm\hat{z}\times (\overline{\overline{\mu}}_t\cdot \bm H_t)&=&\nabla_t E_z\label{gEz}\\
-j\beta \bm H_t-j\omega \bm\hat{z}\times (\overline{\overline{\varepsilon}}_t\cdot \bm E_t)&=&\nabla_t H_z\label{gHz}\\
\bm\hat{z}\cdot(\nabla_t\times\bm E_t)&=&-j\omega \mu_z H_z\label{Hz}\\
\bm\hat{z}\cdot(\nabla_t\times\bm H_t)&=&j\omega \varepsilon_z E_z\label{Ez}
\end{eqnarray}
\end{subequations}
from which $\bm E_t$ and $\bm H_t$ can be easily obtained in terms of $E_z$ and $H_z$. It has been shown that $E_z$ is equal to zero for the case
considered in this paper, therefore the tensorial form of \eqref{eq:EtHt} are
\begin{subequations}
\begin{eqnarray}
H_r&=&\frac{-j\beta}{\omega^2\varepsilon_\varphi \mu_r-\beta^2}\frac{\partial H_z}{\partial r} \label{eq:HrHz}\\
H_\varphi&=&\frac{-j\beta}{\omega^2\varepsilon_r \mu_\varphi-\beta^2}\frac{1}{r}\frac{\partial H_z}{\partial \varphi} \label{eq:HphiHz}\\
E_r &=&-\frac{j\omega\mu_\varphi}{\omega^2\mu_\varphi \varepsilon_r-\beta^2}\frac{1}{r}\frac{\partial H_z}{\partial \varphi} \label{eq:ErHz}\\
E_\varphi&=&\frac{j\omega\mu_r}{\omega^2\mu_r\varepsilon_\varphi-\beta^2}\frac{\partial H_z}{\partial r}\label{eq:EphiHz}
\end{eqnarray}
\end{subequations}
With these relationships and the solution for the dispersion relation given in \eqref{eq:dispersion}, it is possible to deduce the effective tensor describing the 
waveguide in the continuous limit. 

As mentioned before, both $E_r$ and $H_\varphi$ must be zero, which, from \eqref{eq:HphiHz} and \eqref{eq:ErHz} allows to conclude that $\varepsilon_r=\infty$. Additionally,
the dispersion relation \eqref{eq:dispersion} is obtained from the continuity of the $H_z$ and $E_\varphi$ fields, therefore if this solution must be consistent with \eqref{eq:EphiHz}, it is required that $\mu_r=\xi \mu_d$ and $\mu_r\varepsilon_\varphi=\mu_d\varepsilon_d$, which is equivalent to $\varepsilon_\varphi=\varepsilon_d/\xi$.
\par
The solution for the $H_z$ field given by \eqref{eq:solHz} will also hold in the limit $\beta\to 0$, with $k_{td}=\omega \varepsilon_d\mu_d$. In this
case, $H_z$ must be a solution of the two dimensional wave equation in an anisotropic medium, given by \cite{Chew1999}
\begin{equation}
\label{eq:2DHz}
-\frac{1}{r}\frac{\partial}{\partial r}\left(r\frac{\partial H_z}{\partial r}\right)-\frac{\varepsilon_\varphi}{\varepsilon_r}\frac{1}{r^2}\frac{\partial^2H_z}{\partial \varphi^2}=\omega^2\mu_z\varepsilon_\varphi H_z
\end{equation}
whose solutions are linear combinations of cylindrical harmonics and Bessel functions of real order, i.e.
\begin{equation}
H_z(r,\varphi)=\sum_{q=-\infty}^\infty A_q J_{q P}(k r)e^{jq\varphi}
\end{equation}
where $P=\sqrt{\varepsilon_\varphi/\varepsilon_r}$ and $k=\omega\sqrt{\varepsilon_\varphi\mu_z}$. It is straightforward to show that for the above equation be consistent with \eqref{eq:solHz} we need that $\mu_z=\mu_r=\xi \mu_d$. Note that since $\varepsilon_r=\infty$, $P=0$ and the radial dependence of the
field will be proportional to $J_0(\omega\sqrt{\varepsilon_d\mu_d}r)$.

The only remaining tensor components to be determined are $\varepsilon_z$ and $\mu_\varphi$. Since $E_z=0$, from \eqref{Ez} we must conclude that $\varepsilon_z=\infty$, but
none of the equations require any assumption to determine the value of $\mu_\varphi$. However, it can be obtained if a two dimensional scattering process under $E_z$ illumination
is assumed. 

Thus, under these conditions, it is known that the equations for $H_z$ and $E_z$ are decoupled, and, in this case, boundary conditions would only imply continuity of $E_z$ and $H_\varphi$.
The previous analysis has shown that $E_z$ must cancell at the boundary of the cylinder, which is enough to define the scattering problem. However, it has been 
also shown that $H_\varphi=0$, 
and, since these fields are related in the two-dimensional case by
\begin{equation}
 j\omega\mu_\varphi H_\varphi=\frac{\partial E_z}{\partial r}
\end{equation}
if $\mu_\varphi$ is allowed to have a finite value, the above problem would be not well defined, since the scattering problem would require that both $E_z$ and $\partial_r E_z$ be
zero at the surface, which is not possible. Thus, we conclude that $\mu_\varphi=\infty$, which means that $\partial_r E_z$ is different than zero and the problem of scattering
must be solved by imposing that $E_z=0$. 

\begin{figure}
\centering
\includegraphics{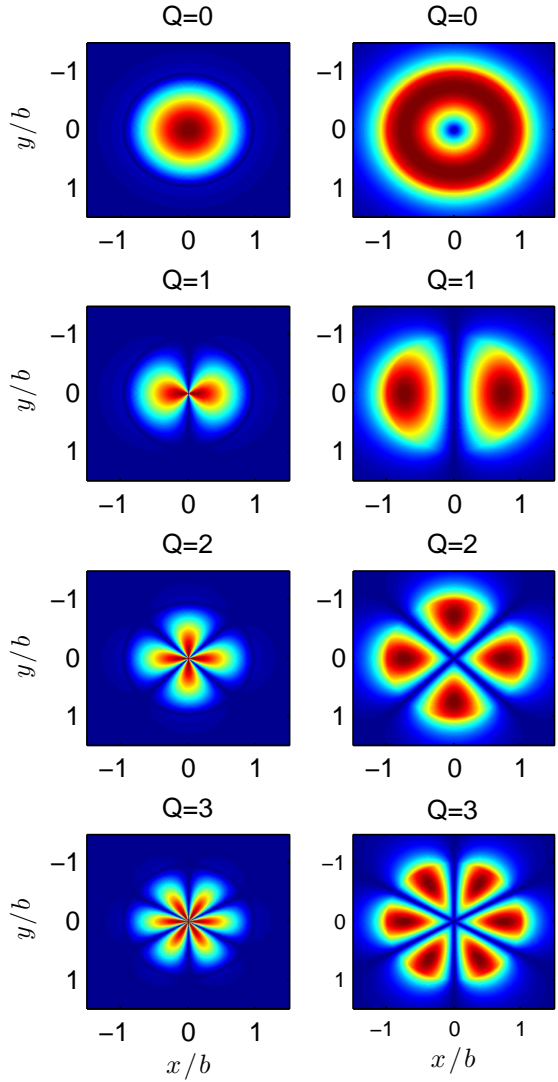}
\caption{\label{figure_modes} Modal solution of the field components $H_z$(left column) and $E_\varphi$ (right column) for the multipolar solutions $q=0,1,2$ and 3. }
\end{figure}

In summary, all the tensor components of the sector cylinder have been determined in the continuous limit, being
\begin{equation}
\overline{\overline{\varepsilon}}=\left(
\begin{array}{ccc}
\infty & 0 & 0 \\
0 & \varepsilon_d/\xi & 0 \\
0 & 0 &\infty \\
\end{array} \right )
\end{equation}
and
\begin{equation}
\overline{\overline{\mu}}=
\left(
\begin{array}{ccc}
\xi\mu_d & 0 & 0 \\
0 & \infty& 0 \\
0 & 0 &\xi\mu_d \\
\end{array} \right ) 
\end{equation}
Once the tensor components have been obtained, the non-zero electric and magnetic field components inside the cylinder can be obtained from \eqref{eq:HrHz} and \eqref{eq:EphiHz}, together with \eqref{eq:solHz}, 
\begin{subequations}
\begin{eqnarray}
H_r(r,\varphi)&=&\frac{j\beta}{k_{td}r}J_1(k_{td}r)\sum_{q=-\infty}^\infty B_q^He^{jq\varphi}\\
E_\varphi(r,\varphi)&=&-\frac{j\omega \xi \mu_d}{k_{td}}J_1(k_{td}r)\sum_{q=-\infty}^\infty B_q^He^{jq\varphi}
\end{eqnarray}
\end{subequations}
Fig. \ref{figure_modes} shows the modal solution of the fields $H_z$ (left column) and $E_\varphi$ (right column) corresponding to $q=0,1,2$ and 3 
for the same waveguide parameters as those considered in the previous section with $\xi=0.1$. The free space wavenumber is $k_0a=\pi$. Notice that all these modes have the same 
radial dependence, proportional to $J_0$ for $H_z$, and to $J_1$ for $E_\varphi$, and that they are also highly degenerated, with $\beta a\approx 4.35$, being the
only difference among them the angular dependence of the EM fields.

\section{CONCLUSIONS}
In this work, the dispersion relation of a waveguide with alternating metallic and homogeneous material sectors has been analyzed in the continuous
limit, that is, when  the number of sectors goes to infinity but the relative filling fraction remains constant. It has been found that this
waveguide allows propagation of only $TE$ modes, since the waveguide effectively behaves as a homogeneous metallic cylinder for the $TM$ polarization. 
\par
In the continuous limit the waveguide becomes azimuthally symmetric, therefore the different multipolar components $q$ are uncoupled each
other. In principle, a different dispersion relation for each multipolar mode is found; however, it has been shown that such modal dispersion is low, 
becoming negligible when the metal filling fraction is increased. The obtained dispersion behavior has been compared with a dielectric rod waveguide, and it has been found that 
these dispersion curves are similar each other, but with the peculiarity that the sectorial waveguide presents negligible modal dispersion.
\par
It has been also demonstrated that in the continuous limit the waveguide behaves as an anisotropic cylinder, and the effective dielectric and magnetic corresponding tensors 
describing it have also be found.

Finally, let us conclude that the main application of these waveguides would be as polarizers, because only the TE polarization is allowed, and the 
dispersion behavior is similar to propagation in free space. Also, the extraordinary anisotropic properties of the rod in the continuous limit suggest its use as
building blocks for new electromagnetic metamaterials.


\section*{Acknowledgment}
This work was partially supported by the Spanish Ministerio
de Ciencia e Innovacion (MICINN) under Contracts No.
TEC2010-19751 and No. CSD2008-66 (the CONSOLIDER
program) and by the US Office of Naval Research.
\ifCLASSOPTIONcaptionsoff
  \newpage
\fi


\end{document}